# "PART MAN, PART MACHINE, ALL COP": AUTOMATION IN POLICING

ANGELIKA ADENSAMER AND LUKAS DANIEL KLAUSNER

ABSTRACT. Digitisation, automation and datafication permeate policing and justice more and more each year – from predictive policing methods through recidivism prediction to automated biometric identification at the border. The sociotechnical issues surrounding the use of such systems raise questions and reveal problems, both old and new. Our article reviews contemporary issues surrounding automation in policing and the legal system, finds common issues and themes in various different examples, introduces the distinction between human "retail bias" and algorithmic "wholesale bias", and argues for shifting the viewpoint on the debate to focus on both workers' rights and organisational responsibility as well as fundamental rights and the right to an effective remedy.

## 1. Digitisation in Policing and Justice

Since the late 20th century, digitisation (transforming information into computer-readable formats), automation (reducing or eliminating the human role in a system or process through the use of computers and algorithms) and datafication (measuring and quantifying people's lives, in particular qualitative aspects thereof, and using the resulting quantified data for various purposes) have hit policing with full force. From predictive policing methods through recidivism prediction to automated biometric identification at the border, more and more aspects of policing employ automated systems.

For example, large databases are algorithmically sifted to detect "suspicious" people and behaviour, potentially leading to stops and searches on the basis of automatic hits. One of these is the database collecting Passenger Name Records (PNR), made compulsory in the EU in 2018 through the EU's PNR Directive (Directive (EU) 2016/681), according to which airlines must transmit PNR of all passengers to police authorities. For example, in the comparatively small country of Austria, 23 877 277 entries have already been recorded – even though in the first 8 months, data collection did not yet include all airlines operating in Austria (Bundeskriminalamt 2019, p. 7). Due to its predictive and explorative nature, the PNR system constitutes an example for the problematic technique of predictive policing (for a more in-depth exemplary discussion of predictive policing in the case of Austria and the United States, see Adensamer and Klausner (2019) and Benbouzid (2019)).

---

Predicting crime is an endeavour currently undertaken by police forces in many countries, seemingly as popular as it is ultimately futile: classic models, like PRECOBS in Germany, KeyCrime and eSecurity in Italy (Alfter et al. 2019, p. 91) or comparable software in Belgium (Alfter et al. 2019, p. 44) and the Netherlands (Alfter et al. 2019, p. 94) are supposed to predict crimes based on historical data of policing and criminal activity. In the field of justice, in a similar vein, the *Correctional Offender Management Profiling for Alternative Sanctions* (COMPAS) algorithm is used in several US states to predict recidivism likelihood of criminal convicts (Larson et al. 2016).

Another currently growing field of digitisation in policing are large-scale video surveillance systems, often with automated face or behaviour recognition. Clearview AI, a product allegedly purchased by several large police forces, recently made the news when it was advertised as containing more than 3 billion images of people scraped from the internet (Hill 2020). Automated (or "assisted") facial recognition has also been implemented and researched in the UK (Fussey, Davies and Innes 2020).

Predictive policing systems are often faced with stark criticism; the use of some tools has been stopped (often after yielding disappointing or inconclusive results, such as PRECOBS in Baden-Württemberg (Mayer 2019)). In Santa Cruz, predictive policing has been banned entirely (Ibarra 2020) over civil liberty and racial discrimination concerns. Similarly, there has been strong resistance against the use of facial recognition as a surveillance tool from civil society, NGOs and activists, and some cities (such as San Francisco (Conger, Fausset and Kovaleski 2019), Boston (Jarmanning 2020) and Portland (Hatmaker 2020)) have already banned the use of facial recognition software, motivated by concerns about racial and gender biases, false positives, privacy and excessive surveillance.

The basis for these kinds of systems very often lies in the creation of massive databases on people and their behaviour. To give just a few examples, the European Dactyloscopy Database (Eurodac) stores over 5 million sets of fingerprints (European Union Agency for the Operational Management of Large-Scale IT Systems in the Area of Freedom, Security and Justice (eu-LISA) 2019, p. 11), mostly of asylum seekers; the Visa Information System (VIS) contains over 27 million registered visa applications with fingerprints (European Union Agency for the Operational Management of Large-Scale IT Systems in the Area of Freedom, Security and Justice (eu-LISA) 2018, p. 28) and the new Entry/Exit System (EES) collects fingerprints of third-country nationals entering the EU (regardless of whether they are visa holders or visa exempt), projected to affect an estimated 295 million people in 2025 (Napieralski 2019, p. 200). Moreover, previously existing databases are increasingly being analysed using novel systems which change the function and effect of these databases entirely. For example, the US Immigration and Customs Enforcement (ICE) used databases containing millions of images from driver's licences for searches with facial recognition software (Harwell and Cox 2020).

While such systems may have their benefits – such as a higher efficiency/throughput rate, the ability to treat comparable cases more uniformly (even in a large and/or distributed organisation), enabling the easier synthesis of data stemming from different sources or the possibility to process standard cases more quickly and allocate more time to more complex cases –, they nonetheless raise several important questions. In this article, we give a comprehensive overview and review of sociotechnical issues in and around automation in policing and the legal systems, find common issues and themes in various different examples, analyse the legal situation in Europe (from the viewpoint of fundamental rights and data protection law) and argue for viewing the debate from a different angle to focus on both workers' rights and organisational responsibility as well as fundamental rights and the right to an effective remedy.

## 2. Problems of Data, Society and Technology

In this section, we give an overview of some of the different ways in which automating processes in policing and justice can be problematic in terms of the underlying data, algorithmic systems and practical, technical and organisational viewpoints. We will first follow the typical data pipeline, going from the input base data through the algorithms and models used to the evaluation and assessment of the results. Then, we will elaborate on the problems caused by scale, in particular the base rate fallacy, and conclude with an argument to consider the debate from the angle of organisational responsibility and the question of workers' rights (i.e. the effect of automation on employees and workers having to work with algorithmic systems) in the context of automation in policing.

2.1. **Flawed and Dirty Data.** The technical and practical side of automation in policing harbours a number of complicated problems beginning at the most fundamental level, the base data themselves. These reflect the imperfect, unequal and discriminatory systems and societies from which they stem (Hao 2019). Following the adage "garbage in, garbage out", any such faulty data will engender mistaken outcomes even if, apart from the data, all other parts of the system were working perfectly (a strong premise, we hasten to add). Any decisions taken on the basis of data have to factor in their provenance and quality and what kinds of distortion they might consequently be subject to, and consider how best to counteract and remediate any such bias – otherwise, any previous unequal treatment or discrimination visible in the data will merely be reproduced or even reinforced (Singelnstein 2018, p. 4).

For example, in the case of predictive policing, prior cases of inferior, discriminatory or outright illegal policing ("dirty policing") are visible in criminological data ("dirty data") (Richardson, Schultz and Crawford 2019, p. 192), e.g. due to underreporting of sexual crimes (Taylor and Gassner 2010, p. 241 ff.), racist crimes (Kushnick 1999, ¶1.7) and police violence (Loftin, McDowall and Xie 2017; Gingerich and Oliveros 2018). Following the legal precepts of non-discrimination in policing and counteracting such biases is made all the harder by the fact that there is often little awareness and acknowledgement of these underlying problems and the discriminatory structures which are (at least partially) responsible therefor. This issue is not helped by the fact that people from marginalised communities are severely underrepresented within the police force (Myers West, Whittaker and Crawford 2019, p. 15 ff.), which tends to correlate with the unequal treatment of minorities by the police and in turn increase the risk of discrimination and unconscious bias (Legewie and Fagan 2016) (though the overall research on the interaction of minority representation in the police and discriminatory policing practices remains inconclusive so far (B. W. Smith 2003; S. Nicholson-Crotty, J. Nicholson-Crotty and Fernandez 2017)).

Similar problems arise in the case of recidivism prediction. ProPublica has criticised anti-Black racial bias (Angwin et al. 2016) in the COMPAS system as used in Broward County, Florida. While ProPublica's reporting has in turn been questioned and criticised (cf. Espino 2018, p. 2 ff.), fundamental problems of fairness (Chouldechova 2017) and transparency (Rudin, Wang and Coker 2020) in recidivism prediction and the underlying data remain, as does the fact that societal racism seems to contribute to unequal rates of recidivism of people of colour beyond what would be expected from hypothesised purely criminogenic risk (Berry et al. 2020).[1]

A joint problem of techniques employing large amounts of data is that even the interpretation of primary data can be problematic. A superficial analysis of the geographical distribution of cases (of whichever crime), e.g., merely focuses the attention towards areas with a higher population density or a similarly trivial reason for higher case incidence; when trying to correct for this, the choice of the basic reference value alone can have a deciding influence on the evaluation – consider the difference in results for neighbourhoods close to train stations or other transport hubs when relating the absolute data to the residential population as opposed to relating them to the ambient population (Belina 2016, p. 92 ff.).

For biometric data, whether fingerprints, DNA or facial images, specific further kinds of problems exist in the base data. Latent fingerprints are often imperfect (Ulery, Hicklin, Kiebuzinski et al. 2013) (distorted, smudged and/or partial) and DNA evidence collected from crime scenes is often contaminated (Fonneløp et al. 2016) (probably most infamously in the case of the so-called "Phantom of Heilbronn", a purported criminal tied to at least 40 crime scenes ranging from burglary and robbery to murder, who turned out to be a worker in a factory producing cotton swabs for investigative uses (Balk 2015, p. 228 f.)). Conversely, due to racial profiling and overpolicing, people of colour are far more likely to be registered in fingerprint (Love 2016) or DNA databases (Krimsky and Simoncelli 2011; Murphy and Tong 2020).[2]

Finally, we turn to the datasets used to train facial recognition algorithms. These have repeatedly been shown to suffer from severe sample bias (Buolamwini and Gebru 2018; Grother, Ngan and Hanaoka 2019), leading to several somewhat unsettling cases of discriminatory results in the recent past. The most notable recent example had Google Photos' automated photo-indexing system label some black people as gorillas, most likely due to black people being underrepresented in the training dataset (Lee 2015). Even three years later, Google had apparently not been able to fix the underlying problem, instead opting to treat the symptoms by blocking labels such as "gorilla" or "chimp" from appearing (Simonite 2018). Moreover, datasets explicitly including race and gender in their annotation often lack critical engagement with and a clear conception and description of the categories involved, severely increasing the likelihood of annotator bias (Scheuerman et al. 2020).[3]

---

[1] In the case of *State v. Loomis*, in which Eric Loomis challenged the use of COMPAS risk assessment in sentencing decision in his criminal case as a violation of his due process rights, the Wisconsin Supreme Court accepted the use of COMPAS in sentencing trials in principle, but warned of its limitations (*Loomis v. State*, Wisconsin Supreme Court, 13. 7. 2016, 881 N.W.2d 749 (Wis. 2016)).

[2] In the case of the EES, the overrepresentation of people of colour in the database is built into the system itself, since only *third-country* nationals' fingerprints are collected, thereby priming the basic population of the database to be disproportionately composed of people of colour.

2.2. **Algorithms and Modelling.** The second set of problems arises from what is actually done with the data. Choosing how to solve a problem, which assumptions to make when modelling a phenomenon and what kind(s) of algorithms to apply to it – all of these involve human decisions prone to implicit or explicit bias. This preexisting bias, to follow the terminology of Friedman and Nissenbaum, is then supplemented and amplified by additional technical and emergent biases (Friedman and Nissenbaum 1996). Much like Paul Watzlawick's famous axiom of communication ("one cannot not communicate"), one cannot not make assumptions. Modelling by necessity means making many decisions representing value systems – which characteristics to interpret positively or negatively, which data values to put into bins together, which characteristics to consider as the opposite ends of a spectrum, etc. Even the choice of not applying any deliberate kind of modelling assumptions, e. g. in the purely correlative and quaintly named predictive policing software HunchLab (Shapiro 2017, p. 459), is a choice. Often, there is little to no deliberate consideration of which simplifications, implicit or explicit assumptions, etc. go into algorithmic or modelling solutions of sociological, societal or even scientific problems (Bennett Moses and Chan 2018, p. 809 ff.). Not least the experiences in modelling the current COVID-19 pandemic have shown with particular clarity that any kind of modelling and prediction needs to value transparency and humility over false decisiveness and conviction in order to "invite insight, not blame" (Saltelli et al. 2020).

To discuss one specific example, the city of Oakland trialled a predictive policing algorithm inspired by seismographic models (Mohler et al. 2015) based on the so-called "near-repeat" theory of criminality. The main computation therein, however, amounts to little more than simply a moving average; it neither accounts for feedback effects on the level of crime (Lum and Isaac 2016, p. 18), nor does it make allowances for the fact that the crime rates of different subgroups of the population might have different elasticities in reaction to increases or decreases in policing – shortcomings which have the potential to entirely reverse the expected results (Harcourt 2007, p. 23 ff.). In general, second and higher order effects (i. e. cascading and feedback effects) have to be an important consideration for any kind of statistically motivated strategy and are often not accounted for (Richardson, Schultz and Crawford 2019, p. 20 ff.).

Ultimately, any kind of technique based on pattern recognition suffers from fundamental faults – such as the implicitly necessary assumption of the regularity of the underlying phenomenon (which can lead to paying less frequent crimes too little attention), the enticement towards fighting symptoms instead of underlying causes,[4] and the fact that it can help to obfuscate discriminatory practices both outwardly and inwardly (Kaufmann, Egbert and Leese 2019, p. 11 ff.).

---

[3] Converse attempts to use biometric identifiers to establish conjectured group characteristics of the person(s) in question, thereby grouping "suspect populations" by genetic relationship, racialised physical characteristics or ethnic or racial identity, seem even more pernicious in this regard (and suffer from the same fundamental human biases described above) (Cole 2018).

As far as the common technophile argument of better decisions through the use of sophisticated technology is considered, even relatively complex models for recidivism prediction such as COMPAS have been shown to yield results no more fair or accurate than the predictions made by humans with limited or no criminal justice expertise (Dressel and Farid 2018). However, there is the additional problem of trading in the "retail bias" of individual human decisions for the "wholesale bias" of subjecting larger groups of people to the same automated decision mechanism (and its according biases). (This distinction has, to the best of our knowledge, not been made in these terms so far.) By way of example, consider a human caseworker with an unconscious bias in favour of people with asymmetrical eyebrows. Assuming a caseload of twelve cases per day and 250 workdays a year, about 3 000 people are subject to this caseworker's individual bias each year, with a very limited areal and temporal impact (and even the conceivable possibility of individuals avoiding this specific caseworker should they feel treated unfairly or have heard about their anti-symmetry bias). If instead an algorithmic system required to be used by all caseworkers exhibits this same bias against people with symmetrical eyebrows, the population of people affected thereby is potentially the entire clientele of the organisation in question (assuming, e.g. thirty locations with an average of ten caseworkers each, up to 900 000 people each year), with no feasible avoidance strategy available to them (for a succinct summary of more ways in which automation is fundamentally different with regard to inequality – e.g. opacity, persistence or universality –, see Eubanks 2018, p. 184–188.).

Even biometric identification, which might at first glance appear to be less susceptible to these kinds of problems, is subject to comparable concerns. Latent fingerprint analysis performed by humans has a significant likelihood of errors (Ulery, Hicklin, Buscaglia et al. 2011; Pacheco, Cerchiai and Stoiloff 2014), and the underlying methodology of fingerprint analysis itself has been called into question (Cole 2005). The arguably more harmful false positives are increasingly likely with automation of fingerprint (or DNA) matching, as matching algorithms are more and more likely to find purported close matches in the growing biometric databases (even more so as previously separate national databases are increasingly being integrated in European and international infrastructures) – with the result that such searches are prone to turn up suspects even if the true perpetrator is not even present in the database (Dror and Mnookin 2010).

The black box nature of the deep learning algorithms used in facial recognition technology raises a whole host of questions regarding transparency, explainability and accountability, with entire conferences (e.g. the ACM Conference on Fairness, Accountability, and Transparency (ACM FAccT) or the AAAI/ACM Conference on AI, Ethics, and Society (AAAI/ACM AIES)) and subfields of the machine learning research community devoted to these issues.[5]

---

[4] As per Goodhart's law: "When a measure becomes a target, it ceases to be a good measure."
[5] See also (G. J. D. Smith 2020) for a consideration on what technological black-box-ness means for our cities and societies at a fundamental level.

2.3. **Evaluation and Assessment.** The third area of concern is the evaluation, interpretation and assessment of the results of algorithm decision support, automated pattern matching or other automation technology. Despite increasing attention to and appreciation of the questions of explainability and transparency of algorithmic systems, even the most well-understood technology remains a black box to most end users, making a critical and self-aware use thereof difficult, if not impossible (Ferguson 2017, p. 1165 ff.). This topic is also fundamentally connected to the question of accountability and responsibility, both on behalf of organisations and individuals; for more detail on this question, see subsection 2.5 below.

Moreover, algorithms and technology are often perceived to be more objective and neutral than humans – a viewpoint that both fails to acknowledge the problems of bias in the underlying data stemming from e. g. past (human) discrimination (be it individual, institutional, systemic) and helps to eschew the necessity of reflecting on and justifying one's actions (cf. Lum and Isaac (2016, p. 18 f.), Bennett Moses and Chan (2018, p. 817 f.) and Shapiro (2017, p. 459)). Biometric evidence in particular suffers from this problem of misplaced and excessive faith in technology and its perceived infallibility (Schklar and Diamond 1999) – for example, the use of DNA evidence in otherwise weak, circumstantial criminal cases significantly increases the likelihood of conviction (Dartnall and Goodman-Delahunty 2006). In reality, the quality of biometric evidence is far from perfect (and sometimes even faked (Giannelli 1997)) and can lead to convictions of innocent people in alarming numbers of cases (Naughton and Tan 2011), few of which are successfully reviewed and overturned (although the positive role DNA evidence can play in overturning wrongful convictions also has to be mentioned (Olney and Bonn 2015).

The issue of misplaced faith in biometric evidence and the need for giving jury members sufficient information to correctly understand the actual strength of the evidence presented has been acknowledged in some jurisdictions, in particular in the case of DNA evidence. However, the long and underscrutinised history of fingerprint evidence has led to paradoxical arrangements in which some jurisdictions *require* the presentation of matching DNA evidence to be accompanied with statistical probabilities while fingerprint evidence is conversely *prohibited* from being presented as anything but categorically certain (Neumann 2012; Neumann, Evett and Skerrett 2012).

Similarly, predictive algorithms such as those used in predictive policing often lack the requisite critical evaluation of their effectiveness and the advantages or disadvantages their use brings with it (cf. Bennett Moses and Chan (2018, p. 815 ff.) as well as Belina (2016, p. 93 f.) for further sources). A recent comprehensive literature review found that so far, there is little empirical evidence either in favour of (i. e. proving the promised benefits exist) or against (i. e. validating the existence of expected drawbacks) their use (Meijer and Wessels 2019). Predictive policing algorithms are generally particularly difficult to evaluate as unfulfilled forecasts can always be attributed to either the effect of acting on the algorithm's prediction (and thus e. g. preventing the forecasted event) or to the algorithm failing to predict accurately. Discriminatory use of predictive policing can lead to overpolicing of areas (possibly erroneously) deemed to be more dangerous. Increased police presence in certain areas can then lead to more arrests in those areas, inducing even more policing of the area and thus creating a positive feedback loop (Adensamer and Klausner 2019, p. 422 f.).

A (perhaps somewhat surprising) challenge of evaluating ADS is that the context of its use has to be taken into account, as organisational and operational factors can make a great difference in its effectiveness. In their ethnographic study of assisted facial recognition used by police in South Wales and London, Fussey, Davies and Innes (2020) found that several seemingly small factors influenced the results: the positioning of the camera, the quality of the photos in the comparison data set ("watchlist"), the expectations of the officers as well as their level of ennui while using the system. These and many other factors have to be taken into account when evaluating an algorithmic support system. Moreover, conflicting interests can create fundamental obstacles in the use of crime prediction systems: When an organisation has to prove that their own predictive system is useful, they also are interested in showing its effectiveness, and therefore showing that the predictions are true. At the same time, it is the innate interest of law enforcement agencies to prevent crime; but after its prevention the effectiveness of the prediction cannot be proven anymore.

2.4. **Base Rate Fallacy and Difficulties of Scale.** The increase in scale of datasets by itself can change the efficacy, drawbacks and dangers of a method significantly, with the use of big data biometrics a stark example for this. Current systems of fingerprint matching have a high reliability, but they are still not perfect. No matching system has a 100 % success rate. Fingerprint identification is nonetheless sometimes still treated as if it were infallible (also cf. the previous section); in the UK, for example, a fingerprint match can shift the burden of proof (i. e. proving that the match is incorrect) on the plaintiff (in this case, an asylum seeker contesting a Dublin transfer) (European Union Agency for Fundamental Rights (FRA) 2018, p. 82).

The great trust in the method of fingerprint matching stems from times when it was used on much smaller datasets, e. g. when comparing fingerprints from a crime scene with those of a limited set of suspects. Today, EU agencies are operating several large scale biometric databases, such as the new Entry/Exit System (EES) collecting fingerprints of third-country nationals entering the EU, expected to affect 295 million people in 2025 (Napieralski 2019, p. 200).

When datasets scale, the number of false positives scales with them. Not taking this into account can mean succumbing to the so-called "base rate fallacy" and consequently to highly overestimating the efficacy of a system. Fingerprints are often thought to be unique, but in a million fingerprints, the fingerprints of some pairs of people will have such a high resemblance that a matching system or an expert will not be able to distinguish them (Dror and Mnookin 2010, p. 55). Purely accidental matches (with severe consequences for the victim of such a mistake) hence become more and more likely the bigger the datasets are. If the failure rate of a fingerprint identification system is e. g. assumed to be 0.1 % (a common industry standard (Jain, Feng and Nandakumar 2010, p. 40)), then in a dataset with a million entries, there will be around 1 000 false matches. If 295 million people's fingerprints are collected (as is projected to be the case for the EU's EES), the number of false positive results of the same matching algorithm will amount to about 295 000 (conversely, achieving acceptable false positive identification rates of 1 % or even 0.1 %, both also common industry standards (Watson et al. 2014, p. xiv), in a database containing dozens of millions of entries requires almost unattainably stringent thresholds for the false match rate (Jain, Feng and Nandakumar 2010, p. 40)).

When comparing very similar fingerprints (which is more likely the bigger the dataset), the standard of accuracy of the matching algorithm consequently also has to be far higher than in smaller sets (Dror and Mnookin 2010, p. 56). Moreover, there have to be better mechanisms and provisions to contest a purported fingerprint match (see subsection 3.3 below) to alleviate cases like these, not least due to the facts that false positives can have significant impact on people's lives (they can e. g. be the deciding factor whether someone is allowed to enter a country) and can lead to lengthy and costly administrative procedures to raise complaints and effect corrections.

2.5. **Organisational Responsibility and Workers' Rights.** Finally, as we transition from sociotechnical and societal issues to legal ramifications, we want to change the focus on the question of responsibility for automated or automation-assisted decisions and the consequences for the employees, in this case police officers, involved in them. The use of such technology is prone to cause conflicts of interests between different levels of organisational hierarchies and affect the work lives of employees lacking agency to have a say in decisions surrounding the use of automation. (We will confine ourselves to commenting on some specific aspects regarding automation in policing and justice here; for a more in-depth investigation of the issue of organisational responsibility in the face of automation, see Adensamer, Gsenger and Klausner (2021). For a discussion of questions surrounding algorithmic control and its contestation between employers and workers, see Kellogg, Valentine and Christin (2020), and for an analysis of different ways in which employers are using algorithms to shift risks from themselves to workers, see Moradi and Levy (2020).)

The central issue is that the decision to use or eschew automation is not made by the people who actually have to employ such technology in their everyday work (cf. Faraj, Pachidi and Sayegh 2018, p. 366 f.). Nonetheless, the introduction of e. g. algorithmic decision support (ADS) changes the expectations placed on their output. Typically, employees can still be held responsible for the decisions they make, but are expected to make them with higher efficiency (Vieth and Wagner 2017, p. 20) and more quickly (Zweig, Fischer and Lischka 2018, p. 15). They often have to endorse or decline automated "suggestions" given by an algorithmic system, irrespective of whether they have the necessary documentation and training to understand it sufficiently. Moreover, as explained above in subsection 2.2, the use of algorithms and automation risks severely aggravating the potential of discriminatory decisions (through what we have named "wholesale bias"), which is of particular salience in the context of policing.

We do not wish to be doomsaying without exception – there are positive examples, such as the child welfare hotline workers assessed in the case study of (De-Arteaga, Fogliato and Chouldechova 2020), which showed that in the right circumstances, trained professionals *can* detect and react appropriately to errors and bias in algorithms. Nonetheless, the comprehensive account by (Kolkman 2020) makes an exhaustive and thorough argument (based on several case studies from the Netherlands and the UK) that transparency and in-depth understanding of algorithmic models may be impossible to achieve even for people working with such models professionally.

In general, employees can be held accountable by their employer when they (illegally) discriminate in their decisions. This situation changes when it is an algorithmic system which discriminates, without the employee (directed to do their work using ADS) sufficiently understanding the model or its effects. In that case, we argue that employees cannot be held accountable, because they neither have power over the use of the algorithm nor the means to check its decisions for discrimination in a meaningful way. Responsibility can never exceed the scope for decision-making.

When ADS is introduced, the power (and with it the responsibility) shifts away from the employee (who has previously made decisions without the use of ADS and had more time for each individual decision). Responsibility is now split between the management or government deciding to introduce ADS, the programmers developing the system and the people tasked with quality control of the model. If, between all these parties, it remains unclear who is practically responsible for discrimination in an individual case, what follows in that situation is what we call a "responsibility vacuum". Hence when introducing ADS in an organisation (and in particular so in the case of police and justice), a lot of attention has to be paid to the decision-makers who are newly "supported" by algorithmic systems. Their job description might change implicitly, but their qualifications and knowledge do not automatically change at the same time or pace, which puts them in a particularly untenable situation if their power over decisions diminishes while their degree of responsibility remains.

Regarding risk-shifting, many of the common observations and critiques do not apply in the specific context of policing – for instance, Moradi and Levy identify four main ways in which risk is reallocated using automated systems: highly flexible staffing and scheduling, a redefinition of what compensable work is, the detection and prevention of loss and fraud, and the incentivisation and exhaustion of productivity. What all of these share is the characteristic that existing inefficiencies within an organisation are not eliminated, but that ADS instead "redistribute[s] the risks and costs of these inefficiencies to workers" (Moradi and Levy 2020, p. 278). The sort of systems Moradi and Levy describe are much less likely to be introduced in public bodies like the police force. Instead, in the police force and similar (quasi-)public bodies, we identify the question of discrimination and the responsibility therefor as the central issue when ADS systems are introduced; particularly in the case of the police, shifting personal responsibility even further away from the individual seems especially worrying in a system in which it is already very difficult to successfully fight discriminatory treatment or effect disciplinary measures against individual members of the police who have been shown to exhibit strongly discriminatory treatment. In both cases (the scenarios investigated by Moradi and Levy as well as our analysis of the effect in the police force), we see a shift of burdens through the introduction of algorithmic systems: in one case, the burden of risk, in the other, the burden of responsibility.

## 3. Legal Considerations

We now turn to legal aspects of automation in policing. Many tools serving as examples of automation and digitisation in policing in this article are used for surveillance purposes and for decision-making based on personal data, which leads to important legal questions on privacy and protection of personal data. In the following we will discuss the impact of such tools on fundamental rights, then turn to questions of data protection and finally discuss the right to effective remedy in the context of automation in policing and justice.

3.1. **Fundamental Rights.** Many aspects of automation in policing and particularly measures of mass surveillance are a threat to the protection of fundamental rights, such as the right to respect for private life (Art. 7 Charter of Fundamental Rights (CFR) and Art. 8 European Convention of Human Rights (ECHR)) and the right to protection of personal data (Art. 8 CFR). Whenever authorities process data about persons these rights are infringed, and unless the measures are proportional, they violate fundamental rights. Such measures must be tested on the grounds of necessity, foreseeability, safeguards, oversight and proportionality (in the narrower sense of the word).

Surveillance measures have to be "in accordance with the law" (*Klass and Others v. Germany*, ECtHR, 6. 9. 1978, 5029/71, para. 58). This can be a problem when law enforcement agencies introduce new technologies without an explicit legal basis. In Austria, for example, the police have started using facial recognition technology on the basis of laws allowing for general video surveillance measures (Bundesministerium für Inneres (BMI) 2019); this could be a violation of the principle of legality.

The European Court of Human Rights (ECtHR) has developed a list of minimum safeguards that have to be specified in any law on secret measures of surveillance: (1) the nature of the offences that may give rise to an interception order; (2) a definition of the categories of people liable to be surveilled; (3) a limit on the duration of surveillance; (4) the procedure to be followed for examining, using and storing the data obtained; (5) the precautions to be taken when communicating the data to other parties; and (6) the circumstances in which data may or must be erased (*Weber and Saravia v. Germany*, ECtHR, 29. 6. 2006, 54934/00, para. 95).

In the case of bulk communications surveillance in the UK brought before the court in the wake of the Snowden revelations, the ECtHR has found that oversight over the measures has to include "the entire selection process, including the selection of bearers for interception and the selection of material for examination by an analyst" (*Big Brother Watch and Others v. the United Kingdom*, ECtHR, 13. 9. 2018, 58170/13, 62322/14 and 24960/15, para. 387). The bulk interception in the UK had been determined to be in violation of the right to privacy in Art. 8 ECHR.

In the light of these judgements, it is clear that all systems of mass processing of personal data have to adhere to some intentionality. An entirely open-ended data mining and machine learning approach to wholesale "big data" surveillance cannot satisfy the criteria of oversight over selectors and search criteria for filtering that the ECtHR has put forward in *Big Brother Watch v. the UK*.

In the jurisdiction of the European Court of Justice (ECJ) on surveillance, the judgements on data retention stand out. In the case of *Digital Rights Ireland/Seitlinger and Others* (ECJ, 8. 4. 2014, C-293/12 and C-594/12, ECLI:EU:C:2014:238), the ECJ has found that the retention of data without a concrete case or investigation is a violation of Art. 7 and Art. 8 CFR. This applies even before the data are accessed and analysed – the retention of mass data itself is an infringement of fundamental rights (*Digital Rights Ireland*, para. 34). It is only justified under a number of criteria which the ECJ has further developed in the cases *Schrems* (ECJ, 6. 10. 2015, C-362/14, ECLI:EU:C:2015:650) and *Tele2 Sverige/Watson and Others* (ECJ, 21. 12. 2016, C-203/15 and C-698/15, ECLI:EU:C:2016:970). The legal measures have to be clear and precise, and the ECJ requires minimum safeguards protecting against abuse and unlawful access to personal data (*Schrems*, para. 91). The ECJ also specifically notes that "the need for such safeguards is all the greater where personal data is subjected to automatic processing" (*Schrems*, para. 91). Furthermore, surveillance measures have to be "strictly necessary" (*Schrems*, para. 91; *Digital Rights Ireland*, para. 52) and must not compromise the "essence" of the fundamental right to respect for private life (*Schrems*, para. 94).

The ECJ also declared that surveillance of electronic communications is only permissible for persons with a link (although "even an indirect or remote one" suffices) to serious criminal offences (*Tele2 Sverige*, para. 105; *Digital Rights Ireland*, para. 57). There have to be at least some objective criteria linking the purpose of data processing to the persons whose data are processed. Similar to the opinion of the ECtHR above, according to the ECJ, open-ended data mining of personal data is a violation of fundamental rights.

3.2. **Data Protection.** The processing of personal data by law enforcement agencies in the EU is regulated by the Data Protection Directive for Police and Criminal Justice Authorities (often shortened to "Police Directive" (PD), Directive (EU) 2016/680), whereas the better known General Data Protection Regulation (GDPR) is largely not applicable to the police. As a directive, the Police Directive had to be transposed into member state law by each individual member state (in Austria, for example, this has been carried out through the Datenschutzgesetz (DSG) in 2018). According to Art. 11 of the directive, the member states have to prohibit the automated processing of personal data when it produces adverse legal effects on the person or significantly affects them otherwise (i. e. nonlegally), or if appropriate safeguards for the rights and freedoms of the affected person are not in place. The minimum such safeguard is the right to obtain a human intervention.

Furthermore, automatic decisions and profiling cannot be based on special categories of personal data, such as data revealing racial or ethnic origin, political opinions, religious or philosophical belief, biometric data, sexual orientation, health status, etc., unless suitable safeguards are in place (Art. 11 para. 2 PD). Profiling (i. e. automated processing of personal data to evaluate certain personal aspects relating to a natural person, as defined in Art. 3 para. 4 PD) that leads to discrimination is absolutely prohibited (Art. 11 para. 3 PD).

For automation in policing, this means that decisions with a significant personal impact (e. g. the identity check or search of a person) cannot be made by software alone, but always have to have a "human in the middle" (also known as "human in the loop"). The human in the middle has to be capable of understanding the automated decision in sufficient detail to be able to exert control in a meaningful way (also see subsection 2.5), which puts strong legal limits on the scope of broad, purely correlating, black box algorithms.

3.3. **Right to Effective Remedy.** The right to an effective remedy (Art. 13 ECHR and Art. 47 CFR) functions as a "right to have rights" of sorts. It is an important safeguard for persons affected by automated decisions; at the same time, effective remedies are scarce in the face of intransparent algorithms and diffusion of responsibility (cf. the problem of the "responsibility vacuum" we identified in subsection 2.5). In the case of biometric matching based on EU regulations on biometric data usage at the borders (see above), for example, effective remedies are lacking as such a complaint (in particular, an effective way to dispute the accuracy of a biometric match) is not explicitly regulated; moreover, drawing such a complaint from data protection law alone poses some challenges.

In data protection law, anyone whose data are stored has a right to rectification of their personal data (Art. 16 GDPR; Art. 16 PD, Art. 52 Regulation (EU) 2017/2226 establishing an Entry/Exit System (EES) ("EES Regulation")). In the context of biometrics, this right must include the correction of a false biometric match. It is straightforward to qualify a fingerprint match as personal data, as it by necessity relates closely to an individual – but if the verification of fingerprints is performed in a way such that the result of the match itself is not stored, a legal claim to rectification becomes impossible. In the EES Regulation, for example, the verification process for fingerprints is described in Art. 23, but documenting the results is not explicitly included. There is no provision to store the result of the verification process in any of the databases described in Art. 14 to Art. 20, and therefore no legal basis for storing such data exists. Considering the principle of data minimisation (Art. 5 (1) (c) GDPR, Art. 4 (1) (c) PD), i. e. the principle of not storing any data unless it is absolutely necessary, this is the correct approach. But as a result, new legal instruments have to be found to comply with the right to an effective remedy regarding biometric matching and similar tools, as such a right cannot be derived from data protection law alone.

4. Conclusion

The increase in automation in policing is a trend as wide-spread as it is concerning. Policing and justice in the 21st century have been shaped by discourses on the use of force, discriminatory behaviour and accelerating digitisation; we have given an extensive overview and review of sociotechnical questions raised by automation in policing and justice and discussed commonalities in varied different examples and contexts. These interrelated debates intersect to raise new issues; in particular, we introduced the distinction between human "retail bias" and algorithmic "wholesale bias" and argued that trading in the former for the latter constitutes a paradigmatic shift in the kind of discrimination that is possible, especially in the light of human propensity towards ascribing technical solutions more objectivity than is warranted.

We have found that further research and appropriate regulations are needed to address particularly the question of organisational responsibility for the use of ADS (and automated decision-making) and related issues of workers' rights. Whenever ADS systems are introduced, it has to be ensured that the employees' responsibility for decisions does not exceed their knowledge and scope for decision-making. At the same time, organisational responsibility must not have any gaps, i. e. there must not be any kind of "responsibility vacuum".

Finally, we have analysed the legal situation with a particular focus on fundamental rights and data protection law. Automated policing measures are subject to several legal restrictions. The case law of the ECtHR as well as the ECJ shows that large-scale open-ended data mining of personal data violates fundamental rights. In terms of EU data protection law, automated decisions have to have a "human in the middle" who can exert control in a meaningful way. In some cases of automation, particularly biometric identification, there is currently no effective remedy for the case of false positives; we have argued that this legislative deficit needs to be resolved urgently.

As we have expounded, the use of this kind of automation is fraught with pitfalls and areas of concern, only some of which can be effectively mitigated. Some tools are better avoided altogether.

## Funding

Both authors were supported by the Digitalisation Fund of the Vienna Chamber of Labour through the project B-01 "Algorithms, Law and Society: Decision Makers Between Algorithmic Guidance and Personal Responsibility". The open access costs were funded by the St. Pölten University of Applied Sciences Publication Fund.

## Acknowledgements

We are grateful to the attendees of the Digital Technologies in Policing and Security Session at the EASST + 4S Joint Meeting 2020 as well as to Sabrina Burtscher, Fabian Fischer, Maximilian Heimstädt, Reinhard Kreissl, Nikolaus Pöchhacker, Philipp Sonderegger and Marlies Temper and the referees, Frédéric Amblard and Woodrow Barfield, for suggesting numerous improvements to both the content and the presentation of this paper.

## Court Cases

*Big Brother Watch and Others v. the United Kingdom*, ECtHR, 13. 9. 2018, 58170/13, 62322/14 and 24960/15
*Digital Rights Ireland Ltd v. Minister for Communications, Marine and Natural*

*Resources and Others/Kärntner Landesregierung and Others*, ECJ, 8. 4. 2014, C-293/12 and C-594/12, ECLI:EU:C:2014:238
*Klass and Others v. Germany*, ECtHR, 6. 9. 1978, 5029/71
*Maximilian Schrems v. Data Protection Commissioner*, ECJ, 6. 10. 2015, C-362/14, ECLI:EU:C:2015:650
*State v. Loomis*, Wisconsin Supreme Court, 13. 7. 2016, 881 N.W.2d 749 (Wis. 2016)
*Tele2 Sverige AB v. Post- och telestyrelsen/Secretary of State for the Home Department v. Tom Watson and Others*, ECJ, 21. 12. 2016, C-203/15 and C-698/15, ECLI:EU:C:2016:970
*Weber and Saravia v. Germany*, ECtHR, 29. 6. 2006, 54934/00

## Laws Cited

Charter of Fundamental Rights of the European Union (CFR)
Directive (EU) 2016/680 of the European Parliament and of the Council of 27 April 2016 on the protection of natural persons with regard to the processing of personal data by competent authorities for the purposes of the prevention, investigation, detection or prosecution of criminal offences or the execution of criminal penalties, and on the free movement of such data, and repealing Council Framework Decision 2008/977/JHA (Data Protection Directive for Police and Criminal Justice Authorities, "Police Directive")
Directive (EU) 2016/681 of the European Parliament and of the Council of 27 April 2016 on the use of passenger name record (PNR) data for the prevention, detection, investigation and prosecution of terrorist offences and serious crime (PNR Directive)
European Convention of Human Rights (ECHR)
Regulation (EU) 2016/679 of the European Parliament and of the Council of 27 April 2016 on the protection of natural persons with regard to the processing of personal data and on the free movement of such data, and repealing Directive 95/46/EC (General Data Protection Regulation, GDPR)
Regulation (EU) 2017/2226 of the European Parliament and of the Council of 30 November 2017 establishing an Entry/Exit System (EES) to register entry and exit data and refusal of entry data of third-country nationals crossing the external borders of the Member States and determining the conditions for access to the EES for law enforcement purposes, and amending the Convention implementing the Schengen Agreement and Regulations (EC) No 767/2008 and (EU) No 1077/2011 ("EES Regulation")

## References

Adensamer, Angelika, Rita Gsenger and Lukas Daniel Klausner (2021). '"Computer Says No": Algorithmic Decision Support and Organisational Responsibility'. In: *[under review]*.
Adensamer, Angelika and Lukas Daniel Klausner (2019). 'Ich weiß, was du nächsten Sommer getan haben wirst: Predictive Policing in Österreich'. In: *juridikum* 3/2019, pp. 419–431. DOI: 10.33196/JURIDIKUM201903041901.

Alfter, Brigitte et al. (2019). *Automating Society. Taking Stock of Automated Decision-Making in the EU*. Berlin: AlgorithmWatch. URL: https://algorithmwatch.org/en/automating-society/.

Angwin, Julia et al. (2016-05-23). 'Machine Bias'. In: *ProPublica*. URL: https://www.propublica.org/article/machine-bias-risk-assessments-in-criminal-sentencing (visited on 2020-07-02).

Balk, Carly (2015). 'Reducing Contamination in Forensic Science'. In: *Themis* 3, pp. 222–239. URL: https://scholarworks.sjsu.edu/themis/vol3/iss1/12.

Belina, Bernd (2016). 'Predictive Policing'. In: *Monatsschr. Kriminol. Strafrechtsreform* 99.2, pp. 85–100. DOI: 10.1515/MKS-2016-990201.

Benbouzid, Bilel (2019). 'To Predict and to Manage. Predictive Policing in the United States'. In: *Big Data Soc.* 6.1, pp. 1–13. DOI: 10.1177/2053951719861703.

Bennett Moses, Lyria and Janet Chan (2018). 'Algorithmic Prediction in Policing: Assumptions, Evaluation, and Accountability'. In: *Policing Soc.* 28.7, pp. 806–822. DOI: 10.1080/10439463.2016.1253695.

Berry, Katie Ropes et al. (2020). 'The Intersectional Effects of Race and Gender on Time to Reincarceration'. In: *Justice Q.* 37.1, pp. 132–160. DOI: 10.1080/07418825.2018.1524508.

Bundeskriminalamt (2019-10-08). *Anfragebeantwortung Fluggastdatenanalyse – Datenschutzrechtliche Aspekte*. URL: https://fragdenstaat.at/anfrage/fluggastdatenanalyse-datenschutzrechtliche-aspekte/4549/anhang/_5790___9030___Anfragebeantwortung-Auskunftspflichtgesetz-Teil_II-20191001_geschwaerzt.pdf (visited on 2020-08-05).

Bundesministerium für Inneres (BMI) (2019-09-25). *Beantwortung einer Anfrage zum Gesichtsfelderkennungssystem nach dem Auskunftspflichtgesetz (BMI-KP1000/0623-II/BK/6.3/2)*. URL: https://fragdenstaat.at/anfrage/gesichtserkennung/4519/anhang/Beantwortung-GFE_AuskPflG_2019_09_21_Ranftl_geschwaerzt.pdf.

Buolamwini, Joy and Timnit Gebru (2018). 'Gender Shades: Intersectional Accuracy Disparities in Commercial Gender Classification'. In: *Proceedings of the 1st Conference on Fairness, Accountability and Transparency*. FAT* '18/PMLR Volume 81. New York, NY: PMLR, pp. 77–91. URL: http://proceedings.mlr.press/v81/buolamwini18a.html.

Chouldechova, Alexandra (2017). 'Fair Prediction with Disparate Impact: A Study of Bias in Recidivism Prediction Instruments'. In: *Big Data* 5.2, pp. 153–163. DOI: 10.1089/BIG.2016.0047.

Cole, Simon A. (2005). 'More than Zero: Accounting for Error in Latent Fingerprint Identification'. In: *J. Crim. Law Criminol.* 95.3, pp. 985–1078. URL: https://scholarlycommons.law.northwestern.edu/jclc/vol95/iss3/10/.

– (2018). 'Individual and Collective Identification in Contemporary Forensics'. In: *BioSocieties, Online First*, pp. 1–26. DOI: 10.1057/s41292-018-0142-z.

Conger, Kate, Richard Fausset and Serge Frank Kovaleski (2019-05-14). 'San Francisco Bans Facial Recognition Technology'. In: *New York Times*. URL: https://www.nytimes.com/2019/05/14/us/facial-recognition-ban-san-francisco.html (visited on 2020-08-06).

Dartnall, Stephanie and Jane Goodman-Delahunty (2006). 'Enhancing Juror Understanding of Probabilistic DNA Evidence'. In: *Aust. J. Forensic Sci.* 38.2, pp. 85–96. DOI: 10.1080/00450610609410635.

De-Arteaga, Maria, Riccardo Fogliato and Alexandra Chouldechova (2020). 'A Case for Humans-in-the-Loop: Decisions in the Presence of Erroneous Algorithmic

Scores'. In: *Proceedings of the 2020 CHI Conference on Human Factors in Computing Systems*. CHI '20. Honolulu, HI: ACM, pp. 1–12. DOI: 10.1145/3313831.3376638.

Dressel, Julia and Hany Farid (2018). 'The Accuracy, Fairness, and Limits of Predicting Recidivism'. In: *Sci. Adv.* 4.1, pp. 1–5. DOI: 10.1126/SCIADV.AAO5580.

Dror, Itiel E. and Jennifer L. Mnookin (2010). 'The Use of Technology in Human Expert Domains: Challenges and Risks Arising from the Use of Automated Fingerprint Identification Systems in Forensic Science'. In: *Law Probab. Risk* 9.1, pp. 47–67. DOI: 10.1093/LPR/MGP031.

Espino, Luis Antonio (2018). 'Racism without a Face: Predictive Statistics in the Criminal Justice System'. senior thesis. Pomona College. URL: http://pages.pomona.edu/~jsh04747/Student%20Theses/LuisEspino18.pdf.

Eubanks, Virginia (2018). *Automating Inequality: How High-Tech Tools Profile, Police and Punish the Poor*. New York, NY: St. Martin's Press. ISBN: 978-1-250-07431-7.

European Union Agency for Fundamental Rights (FRA) (2018). *Under Watchful Eyes: Biometrics, EU IT Systems and Fundamental Rights*. Luxembourg: Publications Office of the European Union. ISBN: 978-92-9491-925-0. DOI: 10.2811/136698.

European Union Agency for the Operational Management of Large-Scale IT Systems in the Area of Freedom, Security and Justice (eu-LISA) (2018). *Technical Reports on the Functioning of VIS as per Article 50(3) of the VIS Regulation and Article 17(3) of the VIS Decision*. Luxembourg: Publications Office of the European Union. ISBN: 978-92-95208-71-1. DOI: 10.2857/935830.

– (2019). *Eurodac – Annual Report 2018*. Luxembourg: Publications Office of the European Union. ISBN: 978-92-95208-82-7. DOI: 10.2857/423772.

Faraj, Samer, Stella Pachidi and Karla Sayegh (2018). 'Working and Organizing in the Age of the Learning Algorithm'. In: *Inf. Organ.* 28.1, pp. 62–70. DOI: 10.1016/J.INFOANDORG.2018.02.005.

Ferguson, Andrew Guthrie (2017). 'Policing Predictive Policing'. In: *Wash. Univ. J. Law Policy* 94.5, pp. 1109–1189. URL: https://openscholarship.wustl.edu/law_lawreview/vol94/iss5/5/.

FonnELøp, Ane Elida et al. (2016). 'Contamination during Criminal Investigation: Detecting Police Contamination and Secondary DNA Transfer from Evidence Bags'. In: *Forensic Sci. Int. Genet.* 23, pp. 121–129. DOI: 10.1016/J.FSIGEN.2016.04.003.

Friedman, Batya and Helen Nissenbaum (1996). 'Bias in Computer Systems'. In: *ACM Trans. Inf. Syst.* 14.3, pp. 330–347. DOI: 10.1145/230538.230561.

Fussey, Pete, Bethan Davies and Martin Innes (2020-10). ''Assisted' Facial Recognition and the Reinvention of Suspicion and Discretion in Digital Policing'. In: *Br. J. Criminol.*, azaa068 (online first). DOI: 10.1093/bjc/azaa068.

Giannelli, Paul C. (1997). 'The Abuse of Scientific Evidence in Criminal Cases: The Need for Independent Crime Laboratories'. In: *Va. J. Soc. Policy Law* 439.4, pp. 439–478. URL: https://scholarlycommons.law.case.edu/faculty_publications/762/.

Gingerich, Daniel W. and Virginia Oliveros (2018). 'Police Violence and the Underreporting of Crime'. In: *Econ. Polit.* 30.1, pp. 78–105. DOI: 10.1111/ECPO.12102.

Grother, Patrick, Mei Ngan and Kayee Hanaoka (2019). *Face Recognition Vendor Test Part 3: Demographic Effects*. Vol. 8280. NIST Interagency/Internal Report. Gaithersburg, MD: National Institute of Standards and Technology. DOI: 10.6028/NIST.IR.8280.

Hao, Karen (2019-02-13). 'Police Across the US are Training Crime-Predicting AIs on Falsified Data'. In: *MIT Technology Review*. URL: https://www.technologyreview.com/s/612957/predictive-policing-algorithms-ai-crime-dirty-data/ (visited on 2019-04-26).

Harcourt, Bernard E. (2007). *Against Prediction: Profiling, Policing, and Punishing in an Actuarial Age*. Chicago, IL: University of Chicago Press. ISBN: 978-0-226-31613-0.

Harwell, Drew and Erin Cox (2020-02-27). 'ICE Has Run Facial-Recognition Searches on Millions of Maryland Drivers'. In: *Washington Post*. URL: https://www.washingtonpost.com/technology/2020/02/26/ice-has-run-facial-recognition-searches-millions-maryland-drivers/ (visited on 2020-07-29).

Hatmaker, Taylor (2020-09-10). 'Portland Passes Expansive City Ban on Facial Recognition Tech'. In: *TechCrunch*. URL: https://techcrunch.com/2020/09/09/facial-recognition-ban-portland-oregon/ (visited on 2020-09-10).

Hill, Kashmir (2020-02-10). 'The Secretive Company That Might End Privacy as We Know It'. In: *New York Times*. URL: https://www.nytimes.com/2020/01/18/technology/clearview-privacy-facial-recognition.html (visited on 2020-07-24).

Ibarra, Nicholas (2020-06-23). 'Santa Cruz Becomes First U.S. City to Approve Ban on Predictive Policing'. In: *Santa Cruz Sentinel*. URL: https://www.santacruzsentinel.com/2020/06/23/santa-cruz-becomes-first-u-s-city-to-approve-ban-on-predictive-policing/ (visited on 2020-08-06).

Jain, Anil Kumar, Jianjiang Feng and Karthik Nandakumar (2010). 'Fingerprint Matching'. In: *Computer* 43.2, pp. 36–44. DOI: 10.1109/MC.2010.38.

Jarmanning, Ally (2020-06-24). 'Boston Lawmakers Vote To Ban Use Of Facial Recognition Technology By The City'. In: *NPR*. URL: https://www.npr.org/sections/live-updates-protests-for-racial-justice/2020/06/24/883107627/boston-lawmakers-vote-to-ban-use-of-facial-recognition-technology-by-the-city (visited on 2020-08-06).

Kaufmann, Mareile, Simon Egbert and Matthias Leese (2019-05). 'Predictive Policing and the Politics of Patterns'. In: *Br. J. Criminol.* 59.3, pp. 674–692. DOI: 10.1093/bjc/azy060.

Kellogg, Katherine C., Melissa A. Valentine and Angèle Christin (2020). 'Algorithms at Work: The New Contested Terrain of Control'. In: *Acad. Manag. Ann* 14.1, pp. 366–410. DOI: 10.5465/ANNALS.2018.0174.

Kolkman, Daan (2020). 'The (In)Credibility of Algorithmic Models to Non-Experts'. In: *Inf. Commun. Soc.*, 1–17 (online first). DOI: 10.1080/1369118X.2020.1761860.

Krimsky, Sheldon and Tania Simoncelli (2011). *Genetic Justice: DNA Data Banks, Criminal Investigations, and Civil Liberties*. New York, NY: Columbia University Press. ISBN: 978-0-231-51780-5. DOI: 10.7312/KRIM14520.

Kushnick, Louis (1999). ''Over Policed and Under Protected': Stephen Lawrence, Institutional and Police Practices'. In: *Sociol. Res. Online* 4.1, pp. 1–11. DOI: 10.5153/SRO.241.

Larson, Jeff et al. (2016-05-23). 'How We Analyzed the COMPAS Recidivism Algorithm'. In: *ProPublica*. URL: https://www.propublica.org/article/how-we-analyzed-the-compas-recidivism-algorithm (visited on 2020-08-05).

Lee, Wendy (2015). 'How Tech's Lack of Diversity Leads to Racist Software'. In: *SFGate*. URL: https://www.sfgate.com/business/article/How-tech-s-lack-of-diversity-leads-to-racist-6398224.php (visited on 2020-07-07).

Legewie, Joscha and Jeffrey Fagan (2016). *Group Threat, Police Officer Diversity and the Deadly Use of Police Force*. 14-512 (Columbia Public Law Research Paper). New York, NY: Columbia Public Law School. SSRN: 2778692.

Loftin, Colin, David McDowall and Min Xie (2017). 'Underreporting of Homicides by Police in the United States, 1976-2013'. In: *Homicide Stud.* 21.2, pp. 159–174. DOI: 10.1177/1088767917693358.

Love, David (2016-04-10). 'For Black People, Fingerprinting Is a Double-Edged Sword, Sweeping Up the Innocent'. In: *Atlanta Black Star*. URL: https://atlantablackstar.com/2016/04/10/for-black-people-fingerprinting-is-a-double-edged-sword-sweeping-up-the-innocent/ (visited on 2020-07-28).

Lum, Kristian and William Isaac (2016-10). 'To Predict and Serve?' In: *Significance* 13.5, pp. 14–19. DOI: 10.1111/J.1740-9713.2016.00960.X.

Mayer, Nils (2019-09-03). 'Strobl entscheidet sich gegen Precobs'. In: *Stuttgarter Nachrichten*. URL: https://www.stuttgarter-nachrichten.de/inhalt.aus-fuer-die-einbruchvorhersage-software-strobl-entscheidet-sich-gegen-precobs.19a18735-9c8f-4f1a-bf1b-80b6a3ad0142.html (visited on 2020-08-06).

Meijer, Albert and Martijn Wessels (2019). 'Predictive Policing: Review of Benefits and Drawbacks'. In: *Int. J. Public Adm.* 42.12, pp. 1031–1039. DOI: 10.1080/01900692.2019.1575664.

Mohler, George Owen et al. (2015). 'Randomized Controlled Field Trials of Predictive Policing'. In: *J. Am. Stat. Assoc.* 110.512, pp. 1399–1411. DOI: 10.1080/01621459.2015.1077710.

Moradi, Pegah and Karen Levy (2020). 'The Future of Work in the Age of AI: Displacement or Risk-Shifting?' In: *The Oxford Handbook of Ethics of AI*. Ed. by Markus Dirk Dubber, Frank Pasquale and Sunit Das. Oxford: Oxford University Press, pp. 271–288. ISBN: 978-0-1900-6739-7. DOI: 10.1093/OxfordHB/9780190067397.013.17.

Murphy, Erin and Jun Tong (2020). 'The Racial Composition of Forensic DNA Databases'. In: *Calif. Law Rev.* 108.6, pp. 1847–1911. DOI: 10.15779/Z381G0HV8M.

Myers West, Sarah, Meredith Whittaker and Kate Crawford (2019). *Discriminating Systems: Gender, Race and Power in AI*. New York, NY: AI Now Institute. URL: https://ainowinstitute.org/discriminatingsystems.pdf.

Napieralski, Antoni (2019). 'Collecting Data at EU Smart Borders: Data Protection Challenges of the New Entry/Exit System'. In: *juridikum* 2/2019, pp. 199–209. DOI: 10.33196/JURIDIKUM201902019901.

Naughton, Michael and Gabe Tan (2011). 'The Need for Caution in the Use of DNA Evidence to Avoid Convicting the Innocent'. In: *Int. J. Evid. Proof* 15.3, pp. 245–257. DOI: 10.1350/IJEP.2011.15.3.380.

Neumann, Cedric (2012). 'Fingerprints at the Crime-Scene: Statistically Certain, or Probable?' In: *Significance* 9.1, pp. 21–25. DOI: 10.1111/J.1740-9713.2012.00539.X.

Neumann, Cedric, Ian W. Evett and James Skerrett (2012). 'Quantifying the Weight of Evidence from a Forensic Fingerprint Comparison: A New Paradigm'. In: *J. R. Stat. Soc. Ser. A Stat. Soc.* 175.2, pp. 371–415. DOI: 10.1111/J.1467-985X.2011.01027.X.

Nicholson-Crotty, Sean, Jill Nicholson-Crotty and Sergio Fernandez (2017). 'Will More Black Cops Matter? Officer Race and Police-Involved Homicides of Black Citizens'. In: *Public Adm. Rev.* 77.2, pp. 206–216. DOI: 10.1111/puar.12734.

Olney, Maeve and Scott Bonn (2015). 'An Exploratory Study of the Legal and Non-Legal Factors Associated With Exoneration for Wrongful Conviction: The Power of DNA Evidence'. In: *Crim. Justice Policy Rev.* 26.4, pp. 400–420. DOI: 10.1177/0887403414521461.

Pacheco, Igor, Brian Cerchiai and Stephanie Stoiloff (2014). *Miami-Dade Research Study for the Reliability of the ACE-V Process: Accuracy and Precision in Latent Fingerprint Examinations.* Vol. Award Number 2010-DN-BX-K268. Final Technical Report. Miami-Dade, FL: Miami-Dade Police Department Forensic Services Bureau. URL: https://www.crime-scene-investigator.net/reliability-of-the-ace-v-process.html.

Richardson, Rashida, Jason Schultz and Kate Crawford (2019). 'Dirty Data, Bad Predictions: How Civil Rights Violations Impact Police Data, Predictive Policing Systems, and Justice'. In: *N. Y. Univ. Law Rev. Online* 94, pp. 192–233. SSRN: 3333423.

Rudin, Cynthia, Caroline Wang and Beau Coker (2020). 'The Age of Secrecy and Unfairness in Recidivism Prediction'. In: *Harv. Data Sci. Rev.* 2.1. DOI: 10.1162/99608f92.6ED64b30.

Saltelli, Andrea et al. (2020). 'Five Ways to Ensure that Models Serve Society: A Manifesto'. In: *Nature* 582, pp. 482–484. DOI: 10.1038/D41586-020-01812-9.

Scheuerman, Morgan Klaus et al. (2020). 'How We've Taught Algorithms to See Identity: Constructing Race and Gender in Image Databases for Facial Analysis'. In: *Proc. ACM Hum.–Comput. Interact.* 4.CSCW1, 58:1–35. DOI: 10.1145/3392866.

Schklar, Jason and Shari Seidman Diamond (1999). 'Juror Reactions to DNA Evidence: Errors and Expectancies'. In: *Law Hum. Behav.* 23.2, pp. 159–184. DOI: 10.1023/A:1022368801333.

Shapiro, Aaron (2017). 'Reform Predictive Policing'. In: *Nature* 541.7638, pp. 458–460. DOI: 10.1038/541458a.

Simonite, Tom (2018-11-01). 'When It Comes to Gorillas, Google Photos Remains Blind'. In: *WIRED.* URL: https://www.wired.com/story/when-it-comes-to-gorillas-google-photos-remains-blind/ (visited on 2020-07-07).

Singelnstein, Tobias (2018-01). 'Predictive Policing: Algorithmenbasierte Straftatprognosen zur vorausschauenden Kriminalintervention'. In: *Neue Z. Strafr.* 38.1, pp. 1–9.

Smith, Bradley W. (2003). 'The Impact of Police Officer Diversity on Police-Caused Homicides'. In: *Policy Stud. J.* 31.2, pp. 147–162. DOI: 10.1111/1541-0072.t01-1-00009.

Smith, Gavin J. D. (2020). 'The Politics of Algorithmic Governance in the Black Box City'. In: *Big Data Soc.* 7.2, pp. 1–10. DOI: 10.1177/2053951720933989.

Taylor, S. Caroline and Leigh Gassner (2010). 'Stemming the Flow: Challenges for Policing Adult Sexual Assault with regard to Attrition Rates and Under-Reporting of Sexual Offences'. In: *Police Pract. Res.* 11.3, pp. 240–255. DOI: 10.1080/15614260902830153.

Ulery, Bradford T., R. Austin Hicklin, JoAnn Buscaglia et al. (2011). 'Accuracy and Reliability of Forensic Latent Fingerprint Decisions'. In: *Proc. Natl. Acad. Sci. U.S.A.* 108.19, pp. 7733–7738. DOI: 10.1073/PNAS.1018707108.

Ulery, Bradford T., R. Austin Hicklin, George I. Kiebuzinski et al. (2013). 'Understanding the Sufficiency of Information for Latent Fingerprint Value Determinations'. In: *Forensic Sci. Int.* 230.1–3, pp. 99–106. DOI: 10.1016/J.FORSCIINT.2013.01.012.

Vieth, Kilian and Ben Wagner (2017). *Teilhabe, ausgerechnet. Wie algorithmische Prozesse Teilhabechancen beeinflussen können*. Vol. #2. Impuls Algorithmenethik. Gütersloh: Bertelsmann Stiftung. DOI: 10.11586/2017027.

Watson, Craig et al. (2014). *Fingerprint Vendor Technology Evaluation*. Vol. 8034. NIST Interagency/Internal Report. Gaithersburg, MD: National Institute of Standards and Technology. DOI: 10.6028/NIST.IR.8034.

Zweig, Katharina Anna, Sarah Fischer and Konrad Lischka (2018). *Wo Maschinen irren können: Fehlerquellen und Verantwortlichkeiten in Prozessen algorithmischer Entscheidungsfindung*. Vol. #4. Impuls Algorithmenethik. Gütersloh: Bertelsmann Stiftung. DOI: 10.11586/2018006.

Vienna Centre for Societal Security, Paulanergasse 4/8, 1040 Wien, Austria

*Email address*: a.adensamer@posteo.net

Institute of IT Security Research, St. Pölten University of Applied Sciences, Matthias-Corvinus-Strasse 15, 3100 St. Pölten, Austria

*Email address*: mail@l17r.eu

*URL*: https://l17r.eu